# The Emergence of Life in the Light of Evolution

Betül Kaçar[1*], Tom A. Williams[2], Laura Eme[3,4], Johann Peter Gogarten[5,6] Patricia Sanchez-Baracaldo[7], Anja Spang[8,9], Frank O. Aylward[10], Michael Travisano[11], Paula V. Welander[12], Julie A. Huber[13], Vaughn S. Cooper[14,15], Paul E. Turner[16], Timothy W. Lyons[17], Andrew D. Ellington[18], Shelley D. Copley[19,20], Eugene V. Koonin[21], Michael Lynch[22*]

[1]Department of Bacteriology, University of Wisconsin–Madison, Madison, WI, USA
[2]Centre for Evolution, Department of Life Sciences, University of Bath, Bath BA2 7AX, UK
[3]Ecologie Systématique Evolution, CNRS, Université Paris-Saclay, AgroParisTech, Gif-sur-Yvette, France
[4]The University of Rhode Island, Kingston, Rhode Island, RI, USA
[5]Department of Molecular and Cell Biology, University of Connecticut, Storrs, CT, USA
[6]Institute for Systems Genomics, University of Connecticut, Storrs, CT, USA
[7]School of Geographical Sciences, University of Bristol, University Road, Bristol BS8 1SS, UK
[8]Royal Netherlands Institute for Sea Research, Department of Marine Microbiology and Biogeochemistry, P.O. Box 59, NL-1790 AB Den Burg, The Netherlands
[9]Institute for Biodiversity and Ecosystem Dynamics (IBED), University of Amsterdam, Amsterdam, The Netherlands.
[10]Department of Biological Sciences, Virginia Tech, Blacksburg, VA, USA
[11]Department of Ecology, Evolution, and Behavior, University of Minnesota, St Paul, MN, USA
[12]Department of Earth Systems Science, Stanford University, Stanford, CA, USA
[13]Department of Marine Chemistry and Geochemistry, Woods Hole Oceanographic Institution, Woods Hole, MA, USA
[14]Department of Microbiology and Molecular Genetics, School of Medicine, University of Pittsburgh, Pittsburgh, PA, USA
[15]Center for Evolutionary Biology and Medicine, University of Pittsburgh, Pittsburgh, PA, USA
[16]Department of Ecology and Evolutionary Biology, Yale University, New Haven, CT 06520, USA
[17]Department of Earth and Planetary Sciences, University of California, Riverside, CA, USA
[18]Department of Molecular Bioscience, University of Texas at Austin, Austin, TX, USA
[19]Department of Molecular, Cellular and Developmental Biology, University of Colorado, Boulder, CO, USA
[20]Cooperative Institute for Research in Environmental Sciences, University of Colorado, Boulder, CO, USA
[21]Computational Biology Branch, Division of Intramural Research, National Library of Medicine, National Institutes of Health, Bethesda, MD, USA
[22]Center for Mechanisms of Evolution, Arizona State University, Phoenix, AZ, USA

Correspondence: bkacar@wisc.edu and lynch@asu.edu

## ABSTRACT

The origin of life is often framed primarily as a chemical problem, yet life's defining feature is evolution. Advances in geochemistry, prebiotic chemistry and molecular biology have suggested diverse scenarios for the emergence of genomes, metabolism and cellular compartments on the early Earth, but most of these models ignore the relevance of a population-genetics perspective. Here, we argue that origin-of-life research must expand from asking simply how life began to exploring how it evolved from pre-biological systems. Synthesizing evidence from comparative genomics, phylogenetics, biochemistry, and geoscience, we emphasize that the last universal common ancestor (LUCA) was already a complex, ecologically adapted population of cells far removed from the starting point of life, implying a deep, pre-LUCA evolutionary history. We highlight how population genetics, ecology, and synthetic biology can constrain origin-of-life scenarios by making explicit the roles of selection, drift, mutation, horizontal gene transfer, parasites and compartmentalization in shaping early communities. Finally, we outline an evolutionary research agenda spanning proto-metabolic autocatalytic networks, protocells, and the emergence of translation and the transition to DNA genomes, such that qualitative models can be formalized through evolution-driven hypotheses testable with theory and laboratory experiments, including those with synthetic cells.

## INTRODUCTION

The origin of life remains one of the most profound scientific enigmas. We know that life evolved relatively soon after the Earth cooled enough to sustain liquid water. Isotopic evidence suggests biological activity around 4 billion years ago[1,2], whereas phylogenetic analyses suggest the possibility of even earlier beginnings (Figure 1).[3,4]

**Figure 1**

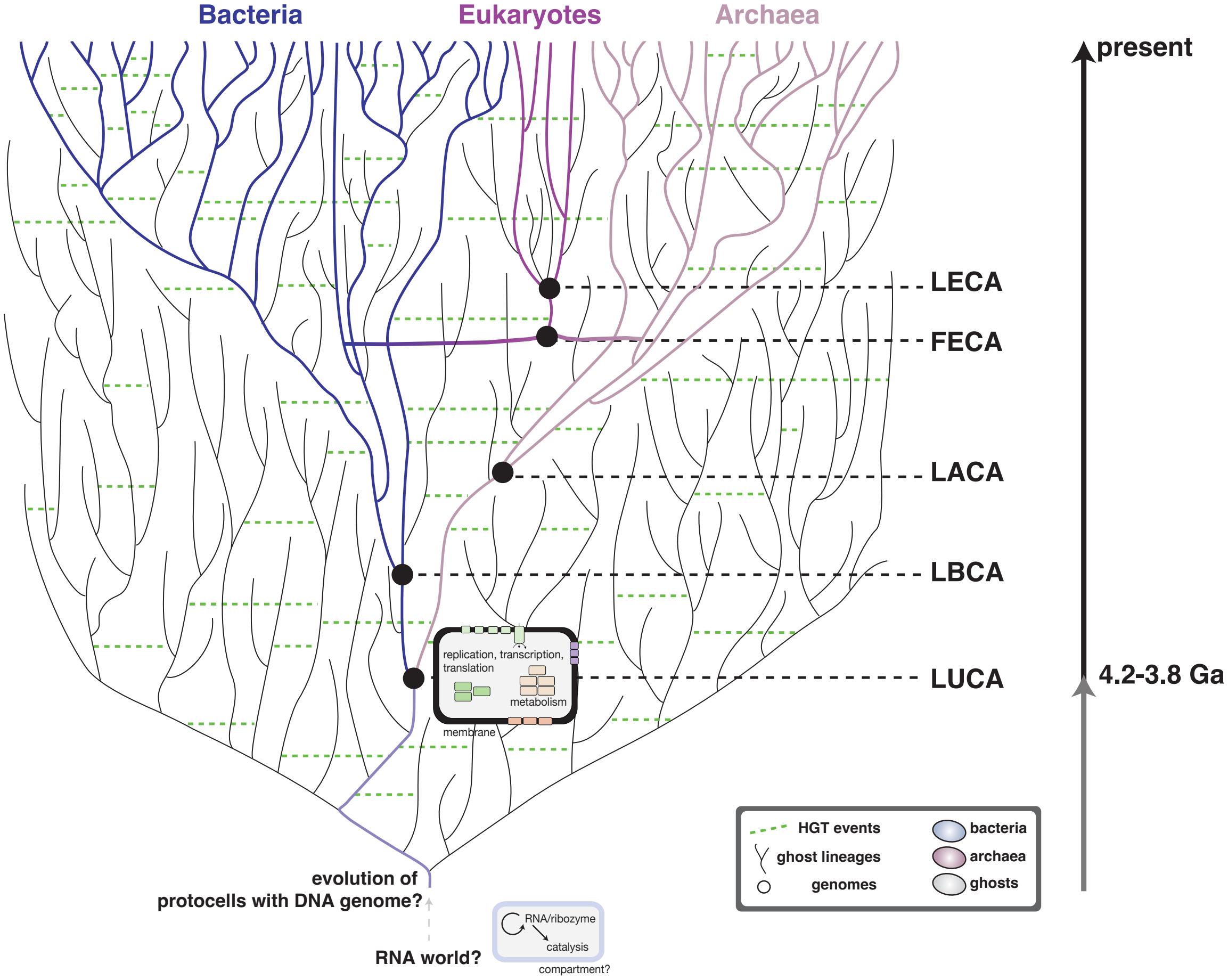

**Figure 1.** A schematic tree of life depicting the evolutionary history of all extant cellular life forms (archaea, bacteria and eukaryotes), building upon a previous scheme[5]. Key ancestors are depicted by black circles: from the last universal common ancestor (LUCA) to the last bacterial and archaeal common ancestors (LBCA and LACA) to the first and last common ancestor of eukaryotes (FECA and LECA). Beside the path leading to extant life, the tree also depicts branches that have left no extant descendants except the genes they may have contributed to the sampled tree of life (green dashed lines denoting horizontal gene

transfer). Although viruses and other genetic elements are not shown for simplicity, they form an integral part of life on Earth and impact the evolution of cellular organisms.

All known life forms share three core features: a genome, a metabolic network, and membrane-bound compartmentalization. The order in which these core features evolved, or perhaps collectively co-evolved, from geological and chemical starting points remains a fascinating open question. Historically, origin-of-life research has been led by geochemistry, chemistry, and biochemistry[6,7]. These fields naturally gravitated toward a "bottom-up" approach, probing how simple molecules could give rise to complex chemical systems. By studying the chemistry of putative prebiotic systems and their progression toward complexity, researchers laid the groundwork for exploring this transition. Specifically, numerous studies on the synthesis of biologically important molecules under various types of conditions deemed to be plausible for the primordial Earth, starting with the classic Miller-Urey experiment[8], demonstrated production of amino acids, nucleobases, sugars and nucleotides[9,10]. Moreover, synthesis of peptides and oligonucleotides under putative primordial conditions has been reported as well, in particular in the presence of mineral surfaces endowed with catalytic capacities[11-13]. More recently, advances in molecular biology, and the resulting ability to draw insights using comparative genomics, phylogenetics, and synthetic biology, have enabled a "top-down" approach, wherein researchers infer early evolutionary trajectories by analyzing modern cellular life and reconstructing its evolutionary history. This agenda has expanded the origins of life field considerably[14,15], providing a data-rich platform for hypothesis testing and model-building[16], with recent advances in both "bottom-up" and "top-down" approaches surveyed in a recent compendium[17]. However, a unified framework to connect early chemical scenarios to the evolutionary principles that drive life's diversification and adaptation remains to be developed.

Despite the impressive progress of both the bottom-up and the top-down approaches, a major conceptual gap remains: how did evolutionary forces, such as natural selection, mutation, genetic drift, horizontal gene transfer (HGT), and other stochastic processes, shape early populations of proto-life forms before the last universal common ancestor (LUCA) (Figure 1)? Although these forces have been universal since life's beginning, they are rarely incorporated into models of early life (but see below for examples of attempts in this direction). Instead, origin-of-life scenarios are often framed as *pre*-evolutionary problems, based on the idea that at least some kind of genetic system would have been required before evolutionary processes as we know them today could begin. However, it is not clear how complex the earliest such systems might have been, with recent theoretical work proposing some forms of evolution taking place in simple, potentially prebiotic settings[18,19]. Therefore, exploring potential scenarios for the origin of life would benefit from closer integration with principles of evolutionary biology to help ensure biological plausibility.

One reason for the disconnect between origin-of-life research and evolutionary theory is historical. The development of tools to probe prebiotic chemistry preceded the rise of molecular biology and especially genomics and phylogenetics. As a result, the field naturally began with a "bottom-up" focus, using the tools of chemistry and biochemistry to investigate the transition from abiotic to biotic systems. In contrast, "top-down" approaches, grounded in comparative genomics, became available much later, as sequencing technologies matured and large-scale genomic data from diverse microbial lineages became accessible. Nevertheless, despite having been established for nearly a century[20,21], the foundations of evolutionary theory have remained conspicuously absent from origin-of-life research.

In our view, origin-of-life research will greatly benefit from a new synthesis treating the emergence of life not just as a chemical puzzle, but also as an evolutionary process[18,22-24]. Such a framework would integrate the constraints and affordances of early geochemical environments with distinct biological boundary conditions, including the constraints on life's emergence and early evolution that were imposed by population structure, (genetic) variation, and selective dynamics[25].

To develop this synthesis, we advocate uniting two underutilized approaches in origin-of-life research: *population genetics*, to understand how evolutionary forces could have shaped early replicators and chemical networks under varying selective regimes, and *synthetic biology*, to experimentally test how such systems behave and evolve under prebiotically plausible constraints. Within this framework, the origin of life would be perceived not as a sequence of molecular inventions arising by unknown mechanisms but as the emergence of evolving systems embedded in ecological and population contexts. Hence 'emergence' in the title of this article, to emphasize that the appearance of life on Earth was, to the best of our understanding, an evolutionary process rather than a singular point of origin, and that life itself is an emergent phenomenon brought about by pre-biological evolution. (Given the long tradition of the 'origin of life' usage, we use the terms 'emergence' and 'origin' interchangeably in what follows). To this end, we provide here a brief overview of the key aspects of the problem, from a plausible RNA World and primordial metabolism to translation, membranes, and DNA, specifically focusing on open questions where approaches from evolutionary and synthetic biology may help the field to progress (Box 1).

**An evolutionary perspective on definitions of life**

Attempts to define life have a long and checkered history, ranging from purely philosophical to operational definitions[26], which we cannot present here in any detail; such a discussion seems to belong more in the field of history and philosophy of science than within the origin of life field as such. The evolutionary perspective suggests recasting the life definition problem that avoids some of the difficulties traditionally associated with this endeavor[14,27,28]. The standard approach is to define life as a phenomenon that possesses some checklist of key traits; for example, the popular NASA definition is that life is "a self-sustaining chemical system capable of Darwinian evolution"[29]. The motivation for, and a major strength of, such a definition is that it provides criteria for searching for life elsewhere in the universe. The downside is that, for any such definition that is specific enough to be useful, it is possible to identify counterexamples among modern life forms on Earth. For example, parasites are obviously alive but not self-sustaining[28]. Indeed, it can be argued that few if any organisms are genuinely self-sustaining, being dependent on trophic, cross-feeding, and other interactions within the communities in which they live. Moreover, the aim of origins research is to understand how life emerged from non-life, at a time when there could not have been much difference between them. In this context, in particular, it is unclear whether prescriptive definitions of life are truly useful. Instead, origins questions relate to the order in which key features of life evolved, and deal with entities that likely possessed some, but not all, of the traits now thought to define life.

There are clear parallels here with similar (but perhaps easier) questions in the study of the more recent evolutionary history. For example, which ancestor of modern birds was the first to merit the label "bird" rather than "dinosaur"? And, which organism on the branch separating eukaryotes from the ancestral archaea qualifies as the first eukaryotic life form [30,31]? Trait-based definitions do not offer non-arbitrary answers to these questions but, as has been recently pointed out[14], phylogenetics provides a helpful framework for doing so through the concept of stem and crown groups [32-34]. Crown groups comprise the common ancestor of the living members of a clade and all its extinct and extant descendants, whereas stem groups include the extinct forms more closely related to the crown than to any other lineage. Thus, crown Life comprises the Last Universal Cellular Ancestor (LUCA) and its descendants, whereas stem Life encompasses lineages both directly ancestral to the LUCA—the steps along the path to modern life considered in traditional origin-of-life scenarios—as well as the (in all likelihood, numerous and diverse) lineages that branched away prior to the LUCA (Figure 1; see [14] for a detailed exposition of this view). This approach recasts life as a clade —albeit an unusual one, with the stem lineage(s) grading out of prebiotic chemistry—rather than a category [28]. Although this view does not resolve where stem life begins, it provides a framework and vocabulary for posing and discussing questions about origins.

**Evolutionary biology of stem, pre-LUCA life**

Without attempting to give yet another trait-based definition of life, it seems useful to outline the attributes that must have coalesced at some point during the stem phase spanning evolution from the emergence of life to the LUCA. We consider that any entity possessing all three core features mentioned above—compartmentalization, metabolism, and information encoding and transmission—would behave, in evolutionary terms, much like life forms as we know them today. The requirement for all three appears to be a straightforward logical necessity. Replicating genomes are required for evolutionary continuity; they cannot persist without a steady supply of monomers produced by the metabolic network, and that network cannot function without sufficiently high concentrations of the substrates that can only be maintained through compartmentalization. Clearly, these criteria accommodate cellular life as we know it, but not viruses or other mobile genetic elements [35,36]. More formally, any living organism is a union of a reproducer, that is, an analogue propagating entity that requires physical continuity (*Omnia cellula e cellula*), and a replicator(s), a digital information carrier (genome) that encodes components of the reproducer. Importantly for the emergence of life discourse, although all three of the above components are essential for life, their material substrates can vary widely and can differ from those of modern cells. Thus, compartmentalization in protocells (hereafter we use this term to denote cell-like entities substantially different from modern cells) might have occurred in vesicles bounded by lipids with a simpler composition than that of the extant cellular membranes or even, initially, in inorganic compartments. Early metabolism likely included simple chemical networks, and some reactions may have differed from modern ones. Genomes could have been small and might have consisted of multiple segments of RNA rather than of DNA. We avoid drawing a sharp boundary between non-living, chemical systems and *bona fide* life within the stem phase of evolution – indeed, that would effectively amount to proposing yet another prescriptive definition.

Instead, consideration of the stem phase within the evolutionary biology framework sets the stage for distinct kinds of questions: not only how the components of life arose, but how early populations, ecosystems, and evolutionary processes shaped the emergence of genomic and cellular complexity. For example, if life during the emergence-to-LUCA period was a consortium, the boundaries between individuals would have been more fluid and the rate of horizontal gene transfer (HGT) would have been substantially higher than in modern systems [37,38]. Such ecological fluidity might have been an important contributor to the emergence of cellular complexity within the relatively short origin-to-LUCA period. As a tantalizing example, phylogenies of some amino-acyl tRNA synthetases seem to bear signatures of HGT from now-extinct lineages.[39] Notably, HGT can enable adaptation of organisms to new environments, including extreme ones[40], which was likely crucial for the early spread of life on earth. However, rampant gene flow is poorly compatible with long-term thriving of organisms under stable environmental conditions and, furthermore, HGT facilitates transmission of genomic parasites[41]. Thus, it is by no means clear that a long phase of rapid gene flow from foreign sources could have been an evolutionarily tractable path to LUCA. More realistically, early evolution would have been shaped, in part, by the trade-off between the beneficial and deleterious effects of HGT, perhaps involving an 'annealing' phase during which the HGT rate was dropping, at different paces, for different classes of genes[37]. Indeed, and as developed further below, the vertical component of evolution from one generation to the next must have been greater than the horizontal component in order to enable selection on individuals and compartments, a consideration that would seem to impose an upper limit to HGT rates even during the earliest stages of evolution. In particular, models of multilevel selection, which is required for the evolution of cooperation and evolutionary transitions in individuality [42-45], would not apply under excessively high HGT rates.

Mutation rates were almost certainly much higher before the evolution of elaborate DNA repair systems, particularly given high levels of radiation on the primordial Earth [46-48], accelerating early molecular evolution, but also imposing severe constraints on the expansion of genome size[49-51]. Hence, another trade-off, between the innovative potential of mutation and the requirement of genetic stability.

Furthermore, a general principle of population genetics is that, for progressive natural selection to become the principal mode of evolution, overwhelming random genetic drift and the accumulation of deleterious mutations, the inequality $N_e s > 1$ (where $N_e$ is the effective population size and $s$ is selection coefficient for beneficial mutations) must hold, and there must be an ample supply of them[52]. Given that, at the earliest stages of evolution, the effective population size of emerging life forms might have been low, drift (that is, chance) could have played an exceptionally important role, with selection being able to operate only on mutations with high fitness effects.

Although we are a bit in the dark on the actual numbers here, the key point is that the principles and main equations of population genetics are completely general and apply to all forms of life, including those that would have been evolving at the earliest stages of evolution on Earth and any that might be discovered on other planets. Therefore, understanding the stem phase of evolution leading to the emergence of the LUCA requires just as much evolutionary-genetic thinking as any problem in modern-day biology.

**Origin of life and major transitions in evolution**

The origin of life, or more precisely, the origin of cells, is the first in the series of major transitions in evolution (MTE), and arguably, the most fundamental and challenging one. Most perceived MTEs involve an evolutionary transition in individuality or, put another way, the emergence of a new Darwinian individual[43,53,54]. Perhaps the most obvious case is the origin of multicellularity, when the emerging Darwinian individual is a collective of cells and selection at the level of individual cells is largely suppressed[55]. There are two substantially different types of MTE: in fraternal transitions, the emerging ensembles unite related individuals, as in the case of multicellularity, whereas in egalitarian transitions, the partners are unrelated, as in the case of the symbiotic origin of eukaryotes[56,57]**.** The emergence of life arguably involved the transition from selection at the level of individual replicators (presumably residing within a protocell or a different type of compartment) to selection of a symbiotic union of an ensemble of replicators and a reproducer (protocell)[58]. As such, the origin of life would qualify as an egalitarian MTE. Under a different angle, however, the origin of life might be considered an autogenic transition[57] because one of the partners (replicators) comes from within the parental system (protocellular compartment).

**The LUCA, a Coalescence Point, Not a Beginning or an End**

The LUCA is a central focus of origin and early life research as it represents the deepest node of inference based on a comparative study of extant life forms (Figure 1). Defined by the universality of the genetic code and the conservation of a core set of about 100 RNA- and protein-coding genes, the LUCA represents the common ancestor of all known cellular life[59]. The sheer antiquity of the LUCA, estimated to date more than 4 billion years ago[4], requires that its characteristics be inferred by tracing back genes and phenotypes found in modern life forms to the roots of the tree of life. Given the vast expanse of time between the LUCA and extant life, it is not surprising that inferences of the gene content of LUCA based on different algorithms and datasets do not always agree.[60] That said, the most recent estimates suggest a genome encompassing over 2,600 genes, similar to the typical complexity of the genomes of modern bacteria and archaea.[4]

Accordingly, the LUCA possessed many hallmarks of modern cells: a fully developed translation system, including ribosomes, tRNAs, aminoacyl-tRNA synthetases and translation factors; transcription and replication systems, although the identity of some key replication enzymes is unclear; a membrane, although its composition remains uncertain[61]; ATP synthase; and a metabolic network capable of producing essential chemical building blocks such as amino acids, nucleotides, and cofactors.[60] The LUCA may have employed the Wood-Ljungdahl pathway for carbon fixation and produced and utilized key cofactors, including NADH, flavins, PLP, and CoA.[62-64]

The inferred sophistication of the LUCA presents a striking puzzle: how did such a complex system arise during the stem phase of evolution that appears to have occurred within a relatively short evolutionary and geological window? Fossil and isotopic evidence places the emergence of tractable life not long after Earth became habitable, suggesting that substantial evolutionary innovation, including the emergence of genomes, metabolism, and membranes, occurred prior to the appearance of the LUCA[5], which itself predates the earliest geochemical evidence of life. That such an advanced form of life emerged so early might suggest that the origin and early evolution of life was facile or even inevitable. To maintain perspective, one should also keep in mind that, although short compared to the overall time of life's evolution on earth, the stem phase still lasted for hundreds of millions of years, an enormous span even if early cell-division times were on the order of months or years.

Despite its antiquity, the LUCA by no account represents the emergence of life, but rather should be considered a waypoint, itself a product of a deeper, complex evolutionary history[4]. The "last" in the LUCA emphasizes that it was preceded by earlier antecedents[37,65,66]: it was not the first life form, but merely the point of coalescence of modern lineages when traced back in time (Figure 1). In that sense, LUCA is analogous to the "mitochondrial Eve", the inferred most recent common ancestor of all extant human mitochondrial DNA lineages, despite having been one individual among many living at the time of coalescence[5]. Just as mitochondrial Eve was not necessarily exceptional among contemporaneous humans, the LUCA was part of a broader community of other prokaryotes, including conspecifics but also many other now-extinct cellular lineages that exchanged genetic material via gene transfer and co-evolved with viruses and other mobile genetic elements[37,67-71]. Put another way, there is no reason to think that the population-genetic processes that shape communities of prokaryotes today were not already in operation at the time of the LUCA. Moreover, detailed phylogenetic analysis of the genes inferred to have been present in the LUCA potentially might uncover traces of genetic exchanges with extinct, ancient lineages.

**Beyond Conceptual Models: From Chemistry to Evolving Populations**

Despite decades of advances, most origin-of-life scenarios remain largely framed in terms of conceptual models anchored in biochemistry and geochemistry. These models typically aim to explain endpoints or prebiotic beginnings, such as the emergence of genetic information, translation, membranes, and metabolism, but usually lack a systematic account of the evolutionary processes that could plausibly connect simpler prebiotic chemical systems to biological complexity exhibited by the LUCA. For example, such a scenario might propose how ribonucleotides formed or how peptides emerged, but leave largely unaddressed the selective pressures, ecological contexts, and competitive dynamics that would have initiated and/or facilitated such innovations and/or allowed their stabilization and spread. Most key features of the life forms that existed in the pre-LUCA period are difficult to infer because, by the very definition of the LUCA, the tree of life antedating it is unobservable, precluding direct reconstruction of evolutionary events. However, other types of evidence can help inferring some of the key steps of the pre-LUCA evolution, in particular the emergence of genomes (probably RNA at the earliest stages), cellularity, metabolism, translation, and the transition to modern DNA genomes. We briefly review these features in the following sections.

Conceptual models tend to treat the emergence of life as a linear progression of molecular innovations, without interrogating the population-level mechanisms that underlie evolutionary change. This lack of population-genetic reasoning, in terms of genetic variation, selection, drift, competition, and cooperation, limits the explanatory power of these accounts. Few studies explore how early replicators would interact, how genetic elements spread or were lost, or how community-level dynamics of protocells could have shaped key transitions[72-74]. Notable exceptions are models of multilevel selection that consider competition at two levels: 1) between replicators within compartments, and 2) between compartments containing ensembles of replicators[75-78]. In particular, the seminal Stochastic Corrector Model of Szathmary and colleagues[75,76] shows how a stochastic ensemble of replicators can overcome the 'Eigen threshold' of the replication error rate, which otherwise is a formidable obstacle to the increase in primordial genome complexity[49,79]. A complementary set of challenges arises from the ecological dimension of origin-of-life models that remains underdeveloped, although various potential crucibles of life have been explored, including hydrothermal vents[80-82], terrestrial geothermal fields[83-85], mineral surfaces[86,87] and others[88]. Here, a promising direction is to adapt paleoclimate models for understanding the functioning of early ecosystems[89]. Moreover, if early life existed in dynamic, interconnected environments, then, from the earliest stages of evolution, selection acted at multiple levels involving molecules, compartments, communities, and not just individuals[18,19,75] (Figure 2). Then, the emergence of life itself should be studied as a complex evolutionary process involving networks of co-evolving entities, such as protocells, and autocatalytic systems, including primordial replicators, where functions emerged and stabilized through variation, drift, selection, mutualism, and parasitism.

## Figure 2

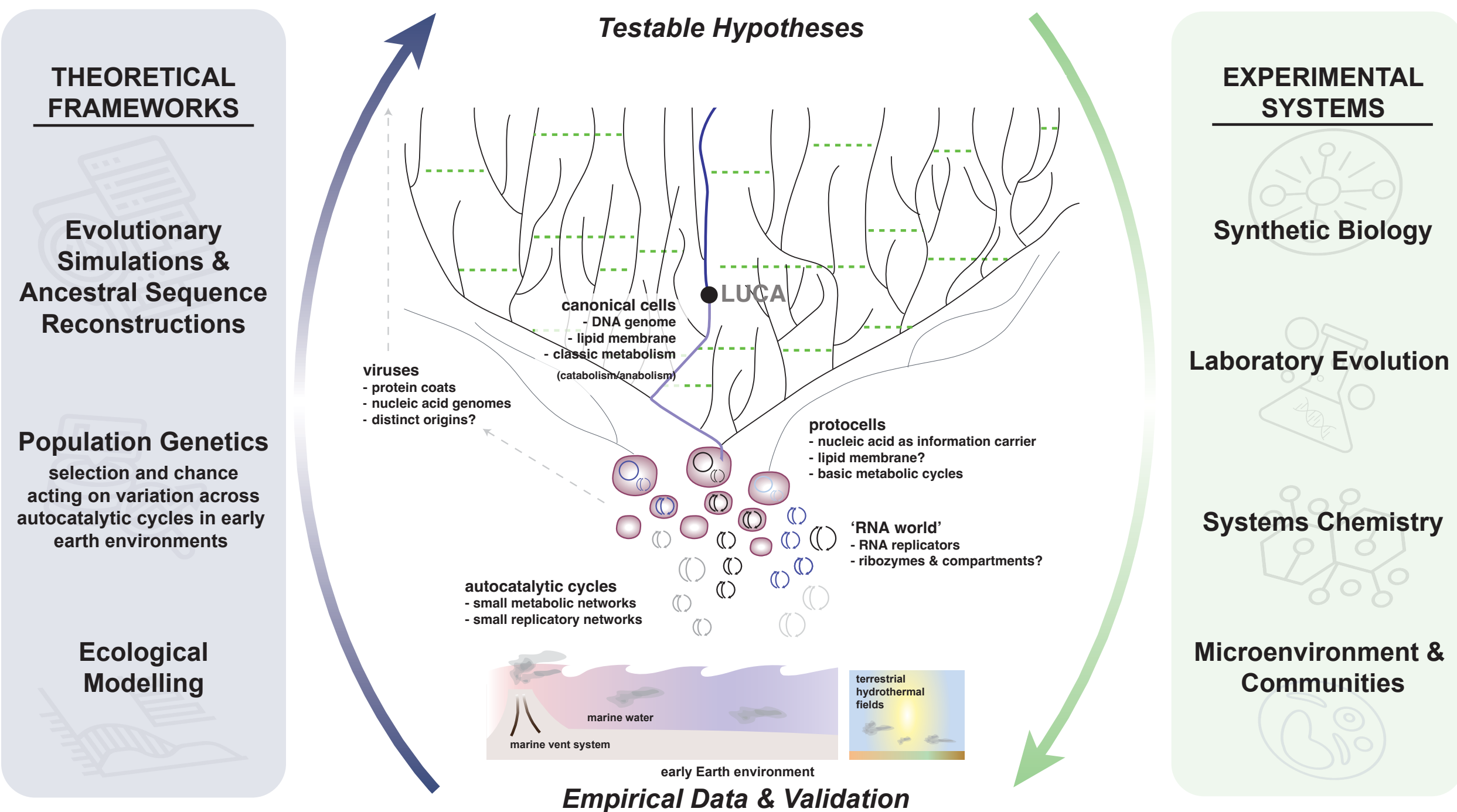

**Figure 2.** Theoretical frameworks and experimental systems promoting progress in our understanding of the early evolution of life (left and right panels, respectively). The central illustration depicts a simplified scenario of pre-cellular evolution involving small autocatalytic cycles (see also [19]) (i.e. metabolic cycles and/or replicators), which may have emerged at various locations across Earth and have interacted symbiotically, including mutualism, commensalism, and parasitism. Some autocatalytic cycles may have acquired physical or chemical membranes (hence compartmentalization), improving their evolutionary potential by enabling concentration of key molecules, information propagation, and continuity across generations—a basis for natural selection. Over time, this process would have allowed the evolution of increasingly more complex interacting cycles, eventually leading to the origin of protocells combining various metabolic and replicatory cycles, that would have ensured more efficient division, replication and enzymatic catalysis, and provided the basis for selection based on genetic variation as well as horizontal gene transfer.

This view, for example, allows us to explore the origin of early polynucleotides or peptides as parasitic or commensal elements, later becoming mutualists as their functions evolved to impose codependence, and possibly to confer benefits such as catalysis or metabolic enhancement. Replicators might have evolved as symbionts (initially, parasites) of protocells whose reproductive success depended on maintaining beneficial internal chemistry, following prescient early ideas[90]. Some mathematical models suggest coordination between replication of emerging genetic elements and protocell division as a key condition of evolutionary stability[58,72,73,91].

This outlook would also recast the origin of the ribosome as a potential case study in functional repurposing. Computational analysis of rRNA suggested an elaborate "onion-skin" model for the evolution of the ribosome, whereby additional sequences accumulated around a small, primordial peptidyl-transferase core that initially lacked the capacity to perform translation but catalyzed non-templated peptide synthesis[92], and chemical specifics of this primordial structure have been suggested.[93,94] However, this model raises two interrelated, fundamental challenges that await empirical evaluation. First, what was the function of the core of the primordial ribosome? Second, how and why was it retained in early organisms long enough for the far more complex function of translation to emerge? A primordial peptidyl-transferase ribozyme might have initially synthesized small, nonspecific peptides.[95] Some of these peptides might have catalyzed beneficial reactions or served as cofactors for ribozymes, increasing protocell fitness and ultimately setting the stage for the evolution of templated translation. Thus, the ribosome might have arisen as a 'catalyst of catalysts' for extant reaction networks that utilized initially random but, later, RNA-encoded peptides.[96,97]

## RNA World and Evolution

A leading hypothesis in the origin-of-life field postulates that an RNA world, in which both informational and catalytic functions were performed by RNA molecules, preceded cellular life as we now know it[98-102]. The RNA-World hypothesis is based in part on the fact that even the modern translation system is composed largely of RNA (i.e., the ribosome and tRNAs), and, moreover, the key peptidyl-transferase reaction is catalyzed by a ribozyme embedded in the ribosomal large subunit RNA[103-106]. Further, there is a striking preponderance of RNA components in cofactors of modern metabolic enzymes (e.g., ATP, NAD, FAD, CoA, folate, and TPP) [107,108]. Recently, following a long history of research on ribozyme RNA polymerases[109-111], the RNA World hypothesis received a major boost from the experimental evolution of a ribozyme of only 45 nt capable of self-replication[112]. Furthermore, various ribozyme reactions, including RNA synthesis and ligation, have been shown to readily occur within fatty acid vesicles, which are plausible protocell models[113-115].

The huge (for a macromolecular complex) size of the ribosome, even in a pared-down form [116], implies that pre-LUCA organisms already produced and replicated large, complex RNA catalysts before the advent of translation. Similarly, it has been argued that cofactors are the oldest surviving 'molecular fossils,' with their adoption by multiple catalysts largely fixing their essentiality to biology (the so-called 'principle of many users'[117]).

Given the chemical complexity that is the pre-requisite to the emergence of metabolic pathways, in particular, those for nucleotide and amino-acid synthesis, and translation, it is likely that at some stage, life was represented by populations of 'ribo-organisms' of cellular-level complexity [107,118-120] that carried multiple RNA genomic segments and made diverse RNA catalysts. This hypothesis is supported by directed-evolution experiments leading to the 'existence proof' of ribozyme catalysts for a wide variety of reactions, including ribozymes that can utilize RNA-based cofactors.[108,121,122] From this vantage point, it seems plausible that at least some riboswitches and ribozymes (catalytic RNAs) in extant life are vestiges of early mechanisms for regulation and catalysis that operated in primordial RNA organisms before the emergence of DNA.[123,124]

Where, how, and even whether the RNA World arose and evolved remain open questions, given the difficulties in prebiotic synthesis for many of its components, the challenge of producing long RNA molecules, and the relative instability of RNA in many environments, particularly, under the harsh conditions of the primordial Earth. At still earlier phases of life's evolution, fundamentally different genetic systems might have preceded RNA. Chemists have explored the possibility of genetic systems using different backbones[125] and different nucleotides.[126-130] To fully realize its explanatory potential, it is important to extend the RNA-World concept to include population-level dynamics, which in part is realized in the models of primordial evolution based on multilevel selection[58,75-78]. Rather than imagining a singular, self-contained ribo-organism, a more realistic picture might include diverse populations of low-fidelity RNA replicators (Figure 2) encoding

RNA catalysts and regulators that were competing, recombining, interacting and evolving within shared environments, but the long-term stability of such systems remains to be explored.

An RNA World in which naked RNA molecules operated as collectives imposes a major problem. The emergence of selfish replicators, that is, genetic parasites—or cheaters, in game-theoretical terms—appears to be part and parcel of the evolution of replicators[72,90]. Indeed, as soon as a resource can be used without producing it evolves, e.g., a replicase that can function *in trans*, cheaters take advantage of it. This argument has been developed formally,[131] and molecular parasites have been shown to emerge in even the simplest models of replicator evolution.[132] Thus, there is every reason to believe that parasitic genetic elements evolved concomitantly with the early replicators. Mathematical modeling of replicator evolution consistently shows that in homogenous, well-mixed environments, parasites take over, leading to eventual collapse of the `entire population[132]. Thus, compartmentalization is a pre-requisite to the emergence of replicator systems capable of sustained evolution (see discussion below and Figure 2). Once a parasite-containing system evolves to the point where the parasite supplies a key resource or function absent from the host and can no longer be lost by the host, becoming a mutualist, and if the parasite is also no longer free to transfer horizontally, permanent linkage of the two partners can set them on an evolutionary course where their interests are fully aligned.

**Membranes and the Architecture of Early Cells**

To enable efficient evolution by natural selection, compartmentalization is required to accumulate and sequester essential molecules, to ensure the continuity between genotype and phenotype, and, as mentioned above, to prevent takeover by parasites. This crucial point follows from the requirement of sufficient concentrations of building blocks for polymerization reactions, but also from the fact that natural selection operates on populations of replicating entities (individuals)[133], and to be effective, requires stable associations between heritable information and function that compartments, however primitive, help maintain. Once such features are in place, evolution is an unavoidable consequence of mutation, selection, and drift[134], and crucially, multilevel selection, at the lower level of individual replicators and at the higher level of compartments, would be in operation from the earliest stages of evolution[58]. Indeed, early population-genetic work suggests that the evolution of compartmentalization would have required both drift and selection: drift to enable genetic differences to emerge between compartments, and selection at the compartment level to promote cooperation between the sets of replicator molecules within[75]. For this reason, and also because compartmentalization is essential for sustaining sufficient concentrations of substrates for RNA synthesis, compartmentalization most likely emerged before genetic systems, rather than afterwards.

In extant bacteria and archaea, membranes consist of lipids with distinct hydrophilic heads and hydrophobic tail regions (amphiphiles). The inferred genome of the LUCA does not encode a full complement of phospholipid synthesis enzymes.[4] Nonetheless, it is virtually certain that the LUCA possessed a cell membrane, given that it encoded the machinery for membrane-protein targeting and insertion (e.g., the signal recognition particle and the Sec system),[135] as well as a membrane-bound ATP-synthase complex essential for energy production.[136-138]

The conundrum is that bacteria and archaea make their membranes from chemically distinct phospholipids synthesized via non-homologous pathways.[61] Bacteria link acyl chains to glycerol head groups via ester linkages to generate sn-glycerol 3-phosphate lipids, whereas archaea connect isoprenoid chains to glycerol head groups via ether linkages to generate sn-glycerol 1-phosphate lipids. These apparently unrelated solutions to the problem of membrane synthesis admit multiple evolutionary possibilities for the membranes of the LUCA [61,139]:

1) LUCA used simple lipids, such as fatty acids, as membrane components, which were subsequently replaced with different complex lipids in bacteria and archaea;

2) LUCA had either the bacterial or the archaeal version of the membranes, with subsequent replacement occurring in the one of the two primary lines of descent. Intriguingly, some recent work suggests that synthetic lipids with archaeal-like features could be more permeable to biomolecules than bacterial lipids, suggesting that archaea-like lipids might have been able to provide a semi-permeable membrane before the evolution of the vast repertoire of membrane transporters found in modern life forms[140];

3) LUCA used both types of phospholipids to produce mixed membranes, with alternative biosynthetic pathways subsequently lost in bacteria or archaea. Notably, mixed lipids are capable of producing protocell-like vesicles[141], and *E. coli* can grow, albeit with a drop in fitness, with a mixture of typical bacterial phospholipids and up to 30% archaeal phospholipids incorporated into the membrane[140,142]

Although the LUCA, in all likelihood, was a population of membrane-bounded cells, membrane vesicles might not have been the first emerging form of pre-biological compartmentalization. In particular, many studies have suggested networks of inorganic compartments existing in the hydrothermal vent walls[80-82] or in terrestrial geothermal fields[83-85] as plausible cradles of life. Indeed, such inorganic compartments provide at least some of the conditions essential for the emergence of life, such as the ability to concentrate organic molecules, energy gradients and catalysts, and potentially could host evolving collectives of primordial replicators[143].

## From Chemical Possibility to Evolutionary Plausibility: Building a Testable Framework

### *Emergence of early metabolism*

By the same token as the early origin of compartmentalization, protometabolic networks providing a steady supply of energy and building blocks for the synthesis of the first replicators must have preceded or at least evolved concomitantly with replication (Figure 2). Modern enzymes are prodigious catalysts, accelerating chemical reactions by up to 26 orders of magnitude.[144] At life's origin, however, such efficacy of catalysis was almost certainly unattainable. Catalysts not only accelerate chemical reactions but play an under-appreciated role in "pruning" complex networks by directing chemical fluxes along specific trajectories at the expense of competing reactions, allowing production of higher levels of fewer components.[145] Early inefficient catalysts, whether they were minerals (e.g., iron sulfides or clays), short peptides, or RNA, likely played a pivotal role during the evolution of rudimentary protometabolic networks by directing fluxes in specific directions.[146]

As protometabolic networks emerged from inorganic precursors, small molecules, such as organic acids, amino acids, and peptides, might have further enhanced catalytic diversity. The reactions required for the synthesis of amino acids, nucleobases, and cofactors are thermodynamically favorable under non-equilibrium conditions, such as those found in and around serpentinizing hydrothermal vents[147]. Indeed, networks of microscopic inorganic compartments, sometimes picturesquely called ‘chemical gardens’, seem to be among the most plausible currently identified cradles of life [80,81,143,148,149]. Notably, it has been shown that addition of a fatty alcohol, decanol, to such structures results in its incorporation into the compartments and vesicle formation, suggesting a path from inorganic to organic compartmentalization[150]. Theoretical studies suggest that, starting from inorganic precursors and a foundational set of small organic compounds, reaction networks could have evolved iteratively, generating the key intermediates of modern metabolisms.[151,152]

From this vantage point, modern metabolic pathways likely reflect two evolutionary trajectories: some descended directly from primordial abiotic networks, with flux gradually being enhanced by ever-more-efficient catalysts, whereas others evolved through pathways explored and stabilized by emergent ribozymes that were subsequently displaced by protein enzymes. Both scenarios would have been shaped by evolving populations of genetically variable protocells across diverse environments, selecting, retaining, and combining metabolic innovations in response to ecological pressures. However, the transition to such complex processes must have involved numerous competing changes, probably during a phase of high mutation rates and therefore not necessarily accumulating in a stepwise fashion. Even a crude depiction of the likelihoods of alternative scenarios will require the implementation of population-genetic models focused on modes of evolutionary divergence. Although this is a highly challenging enterprise, prior research developed in other contexts provides a framework for future progress in this area by suggesting ways to link general evolutionary genetic models to specific molecular and cellular features.[52]

*Emergence of the translation machinery: from symbiotic replicators to the central cellular functionality?*

Building on prior discussions of the RNA World and the possible parasitic origins of catalytic RNAs, recent work has shed new light on the ribosome’s structural evolution and its transition to a central role in translation[153-158]. The ribosome was one of the key evolutionary advancements of early cells, eventually enabling the production of genetically encoded proteins involved in genome replication, energy capture, transport, and catalysis. Clearly, these proteins underpin the organizational and functional complexity of modern life, and the translation system itself is the central functional hub and the main energy consumer in all modern cells[159]. Translation might not have been the original function of the protoribosome. An intriguing possibility is that the ribosome

—or more precisely, proto-ribosomal RNA—was initially a selfish genetic element, a commensal or a parasite of the protocell that either possessed a replicase activity or depended on a replicase encoded by other symbiotic genetic elements[96,97]. In this scenario, once peptide bond formation emerged as a selectable function and some peptides proved beneficial, acting as catalysts or cofactors, the protoribosome would have become a mutualist, paving the way for the evolution of translation. Notably, under a conceptually similar scenario, the eukaryotic mitochondrion might have started as an energy parasite, but eventually became an essential, integral component of the eukaryotic cell once both the host and the symbiont had relinquished key functions.[160] Indeed, recent experiments show that minimal rRNA constructs (~67 nucleotides) can dimerize and catalyze peptide bond formation[95], compatible with the possibility that the protoribosome was a relatively simple RNA molecule that was replicated *in trans* by other, autonomous replicators. Over evolutionary time, such capabilities, initially parasitic or commensal, could have become mutualistic, as peptides improved replication, membrane permeability, or metabolic rates. The transition to templated translation would then represent a major evolutionary transition, marking a shift from analog, collectively catalytic systems to digitally encoded biological information.

Phylogenetic analysis of key proteins involved in translation, such as initiation factors, elongation factors and aminoacyl-tRNA synthetases, shows that they emerged relatively late within their respective protein families.[161,162] The rise of these translation-system components likely coincided with the transition from a ribozyme-centered translation system that achieved sufficient efficiency and fidelity to enable "protein takeover" to the modern, protein-centered system. To better resolve the nature of the earliest ribosome, investigation of the functionality of rRNAs, starting with the minimal peptidyl-transferase ribozyme and further exploring the origins of initiation and elongation factors, could build on recent ancestral sequence reconstructions that suggest generalist functions for the ancestral forms of these factors.[39,163-165] Again, although chemically and biologically plausible, these conceptual models would benefit from the development of formal theory capable of identifying the essential evolutionary conditions that enabled such transitions from parasitism or commensalism to mutualism.[52,72]

*Origin of DNA genomes*

Assuming that the LUCA possessed a genome with many hundreds of genes, the transition from RNA to DNA as the primary genetic material must have occurred at some point during the pre-LUCA evolution. The adoption of DNA as the information carrier likely conferred multiple advantages. In principle, the greater stability of DNA compared to RNA may have enabled the evolution of larger genomes, whereas the allocation of template (DNA) and catalytic (RNA, protein) functions to different types of molecules might have made the system less susceptible to the evolution of genetic parasites.[166]

However, the timing of the evolution of DNA replication remains problematic. Although archaea and bacteria both have DNA genomes, the key enzymes of DNA replication are non-homologous, complicating attempts to reconstruct the ancestral replication machinery.[118] One hypothesis posits that DNA replication and transcription co-evolved from RNA replication systems, possibly through shared polymerase ancestors, prompted by the homology between archaeal replicative DNA polymerase (PolD) and the catalytic subunit of the universal DNA-directed RNA polymerase.[167] Under this scenario, the LUCA already possessed a DNA genome, and the modern, alternative replication machineries were acquired or replaced in different lineages via horizontal gene transfer between different domains of cellular life with the possible involvement of mobile genetic elements.[168]

## Opportunities for Biologists: Combining Experiment with Theory

Despite the uncertainty in many details, it is clear that the evolution of life before LUCA was shaped by diverse populations, ecological interactions, and multiple solutions to fundamental biological problems. The ongoing reframing of the origin of life problem in this context opens many opportunities for evolutionary, population, and synthetic biologists (Figure 2). Looking backwards from modern life, a major impediment to understanding early life is the lack of relevant experimental systems for modeling primordial biology. Synthetic biologists have already generated minimal cells, some with as few as 473 genes[169], but these were constructed by a top-down approach, that is, by deleting non-essential genes from the small genomes of mycoplasmas, and are therefore of limited relevance for understanding the origin of cells. A bottom-up strategy involving assembly of genomes from a reconstructed minimal set of essential genes would be more informative in the context of the origin-of-life research. Very recently, a preliminary report of a striking advance in this direction has been reported, featuring a synthetic cell containing a 90 kb genome assembled from individual genes and split into 8 plasmids, a configuration tantalizingly resembling putative primitive cells from the stem phase of evolution[170]. That minimal cell has been shown to be capable of genome replication, division and selection, providing a chassis for a variety of experiments some of which could be of direct relevance for the emergence of life.

Acknowledging that extant life resulted from a series of singular evolutionary events, a key advance in synthetic biology would be to develop radically alternative, artificial minimal cells, evolutionary plausible, tractable systems, unconstrained by the historical roots of modern cells. A selectively neutral 'ratcheting' mechanism might have contributed to the evolution of protocells, and ultimately, the LUCA, from simpler starting points.[171,172] Clarifying the boundary between chance and necessity in the emergence of key features (to borrow Monod's classic title [173]), such as catalytic RNAs, specific cofactors, ATP synthase, and the proton-motive force, will be crucial to understanding protocell evolution and the origin of life. Construction and evolution of experimental communities of minimal artificial cells could yield breakthrough opportunities for both evolutionary biology and synthetic biology. Such systems need not reproduce early life in detail to be useful. Their value for the origin-of-life research lies in identifying which transitions are evolutionarily stable under specified constraints: for example, whether compartments are required to prevent parasite takeover, whether replicator propagation must be coupled to protocell division, what error thresholds permit the maintenance of functional information, and whether primitive energy-transducing and/or metabolic modules can enhance persistence under fluctuating conditions. Mathematical models and simulations of reproducer–replicator coevolution and parasitism can guide experiments [58,72,73,91], provided that such analyses account for the plausible features of the earliest life forms such as high mutation rates, random genetic drift, and alternative modes of recombination and gene flow.

Equally important will be a deeper investigation into the role of selfish, mobile genetic elements. Although such elements are often viewed solely as parasites, they are vehicles of horizontal gene transfer in the modern biosphere, and the source of many innovations in cellular organisms. Mobile elements likely were essential for transfer of genes among early cells, contributing, in particular, to the transition from RNA to DNA genomes[174], and evolution of DNA replication and transcription machineries.[167]

Moving backwards, at the biochemical level, it is crucial to match the limits of early catalytic capacity—especially in RNA-cofactor systems—against the informational limits imposed by high error rates in early cells. Substantial efforts in this direction have been undertaken, starting with the seminal early work of Eigen and Schuster[49,50,175,176]. More realism can be injected into the study of life's emergence through the development of theoretical models and experimental explorations that explicitly incorporate population size and structure, high mutation rates, horizontal gene transfer, ecological interactions and the physical constraints imposed by compartmentalization and spatial organization. Further, the ways in which cooperation, competition, and evolution can play out in tightly interconnected communities of genetically distinct but metabolically codependent pre-LUCA individuals are squarely within the domain of population biology.

Much remains unknown about the emergence of metabolism as well. Theoretical biologists have long been interested in autocatalytic reaction networks[177-179] that might have preceded the emergence of genetic systems in prebiotic evolution (Figure 2). However, although potential primordial catalysts for key reactions of modern metabolism have been identified,[180-184] there is little empirical evidence of autocatalytic networks capable of sustaining growth and continuity. What is missing, for each key reaction, is a succession of catalysts of increasing specificity and efficiency, starting from abiotic reactions and proceeding through RNA molecules with or without catalytic partners, and eventually leading to genetically encoded proteins, possibly composed of a limited set of early amino acids. A greater challenge is to move beyond catalysis of individual reactions to (proto)metabolic pathways and networks. Advancing from theory to experimental evidence[152] for the function of a minimal autocatalytic reaction network in the presence of primitive catalysts will be an important step forward. Several recent studies have reported notable progress in the experimental reconstruction of protometabolic networks[185-187].

## CONCLUSION

For the progress of the origin of life field, it is crucial to complement chemistry-based research asking how life might have begun with models and analysis rooted in evolutionary biology evaluating the plausibility of alternative hypotheses. We emphasize the principle of continuity, not only in the chemistry of life, but in the evolutionary processes that shape it. The chemical and evolutionary perspectives are highly synergistic: evolutionary biology can inform and provide insights into constraints on the evolutionary pathways explored by early life, whereas chemistry can suggest starting points and routes to early biomolecules that are not obvious from the standpoint of modern biology. To make progress, the field must move beyond conceptual models and embrace experimentally grounded, evolutionarily informed hypotheses made possible by models and tools from population genetics, synthetic biology, and systems modeling (Box 1).

Life's origins should be framed not only as a chemical transition, but also as an evolutionary process, one shaped by ecological structure, driven by population dynamics, and modulated by (genetic) variation, selection, drift, cooperation, and conflict. The LUCA was neither the beginning, nor an ending, but the coalescence point of the modern lineages of life and itself a product of a deep evolutionary history. Understanding that history requires treating the emergence of life not merely as a chemical event, but as the emergence of evolving and ecologically embedded, persistent populations. If we aim to explain life's origin, we must thus treat evolution not as a late consequence but as a defining feature of the process from the outset.

**ACKNOWLEDGMENTS**

We acknowledge the American Academy of Microbiology and the Gordon and Betty Moore Foundation for supporting the Early Microbial Life Colloquium, which provided a foundation for the ideas discussed in this article. We especially thank Nguyen K. Nguyen and Donalyn Scheuner for their leadership in organizing the colloquium.

**BOX 1 | AN EVOLUTIONARY RESEARCH AGENDA FOR THE ORIGIN OF LIFE**

1. **From chemical possibility to evolutionary plausibility**
Develop models that link prebiotic reaction networks to evolving populations of replicators and protocells, explicitly incorporating selection, drift, mutation, recombination and horizontal gene transfer. Use these models to identify which chemically feasible scenarios are evolutionarily stable.

**2. Coevolution of reproducers and replicators, and mutualists and parasites**
Formalize and test theories on how protocellular reproducers and genetic replicators coevolve, including the inevitable emergence of parasites and mutualists. Determine the conditions under which coordination of genome replication and compartment division becomes evolutionarily stable.

**3. Compartmentalization, population structure and ecology**
Investigate how physical compartments (vesicles, pores, mineral matrices, emulsions) create population structure, enable local selection and limit parasite takeover. Combine palaeoenvironmental reconstructions with ecological and population-genetic models of early communities.

**4. The RNA world as a system of evolving populations**
Move beyond a singular "ribo-organism" to study diverse, low-fidelity populations of RNA replicators. Examine how cooperation, competition, recombination and spatial structure influence the emergence and maintenance of catalytic ribozymes and regulatory RNAs.

**5. Origin and evolution of translation**
Treat the ribosome and associated factors as evolving systems. Use ancestral sequence reconstruction, minimal rRNA constructs and synthetic biology to test scenarios for the transition from non-templated peptide synthesis to templated translation and the establishment of the genetic code.

**6. Transition from RNA to DNA genomes**
Model and experimentally probe how division of labor between information storage (DNA) and catalysis (RNA and proteins) could emerge from RNA-based systems. Explore the role of mobile genetic elements and viruses in driving the origin and diversification of DNA replication and transcription machineries.

**7. Emergence and evolution of protometabolism**
Link thermodynamically plausible prebiotic networks of chemical reactions to evolving protocell populations. Reconstruct the evolutionary succession of catalysts, from minerals and short peptides to ribozymes and proteins, that can sustain autocatalytic growth, and test how metabolic innovations spread and persist in structured communities.

**8. Synthetic, minimal and alternative cells as testbeds**
Build and evolve artificial cells and minimal genomes that embody alternative solutions to core biological problems (replication, energy transduction, compartmentalization). Use experimental evolution to distinguish chance from necessity in the emergence of key features such as ATP synthase, proton-motive force and membrane architectures. More generally, synthetic-cell experiments can generate explicit predictions about the minimal conditions under which replication, compartmentalization, energy transduction, and metabolic coupling become jointly sustainable and evolvable.

**Caps need to be removed from some citations.**